*Article*

# Analyzing Factors Associated with Fatal Road Crashes: A Machine Learning Approach

**Ali J. Ghandour [1],\*, Huda Hammoud [2] and Samar Al-Hajj [3]**

[1] National Council for Scientific Research (CNRS), 11-8281 Beirut, Lebanon
[2] Faculty of Engineering and Architecture, American University of Beirut, 1072020, Beirut, Lebanon; hah57@mail.aub.edu
[3] Faculty of Health Sciences, American University of Beirut, 1072020, Beirut, Lebanon; sh137@aub.edu.lb
\* Correspondence: aghandour@cnrs.edu.lb

**Abstract:** Road traffic injury accounts for a substantial human and economic burden globally. Understanding risk factors contributing to fatal injuries is of paramount importance. In this study, we proposed a model that adopts a hybrid ensemble machine learning classifier structured from sequential minimal optimization and decision trees to identify risk factors contributing to fatal road injuries. The model was constructed, trained, tested, and validated using the Lebanese Road Accidents Platform (LRAP) database of 8482 road crash incidents, with fatality occurrence as the outcome variable. A sensitivity analysis was conducted to examine the influence of multiple factors on fatality occurrence. Seven out of the nine selected independent variables were significantly associated with fatality occurrence, namely, crash type, injury severity, spatial cluster-ID, and crash time (hour). Evidence gained from the model data analysis will be adopted by policymakers and key stakeholders to gain insights into major contributing factors associated with fatal road crashes and to translate knowledge into safety programs and enhanced road policies.

## 1. Introduction

Road traffic injury (RTI) imposes a substantial human and economic burden on countries worldwide. More than 1.5 million road users died in 2018 and an additional 50 million were injured or permanently disabled, particularly among vulnerable road users (i.e., pedestrians and motorcyclists) [1]. Addressing the RTI global epidemic is of utmost importance to reduce RTI's associated morbidity and mortality. Understanding key contributing factors that increase the likelihood of sustaining severe and fatal road injury is critical to curtailing the impact of this major public health problem. Existing literature examined major intrinsic and extrinsic factors associated with the increased risk of fatal road crashes. Rolison et al. (2018) assessed the human factors and identified drivers' characteristics (e.g., age, gender, safety measures adopted, risk-taking behavior) that influence the severity of crash outcomes [2]. Other studies examined the role of environmental factors (e.g., road types, nighttime travel, weather conditions) in amplifying the risk of exposure to fatal road injuries. Altwaijri et al. (2011) investigated additional contributing factors, including excessive speed, single vehicle, wet surface, and dark lighting road conditions [3].

Ample research has explored the use of novel methodologies and statistical models to analyze and understand key elements influencing injury severity outcomes and the increased risk of fatal injuries. Statistical and regression models, in addition to feature-based artificial intelligent models, present a non-traditional alternative and technique employed to examine the multiple variables that catalyze the occurrence of road fatalities. These models have been adopted to investigate underlying factors that can contribute to preventing or reducing severe and fatal RTI and guide injury preventive efforts. Keall et al. (2004) and Zajac and Ivan (2002) employed the logit-based model to detect the impact of drinking and driving on the injury severity outcome [4,5]. Lack of safety measures and restraint system use (e.g., seat belt and helmet use) were also identified by Bedard et al. (2002) and Valent et al. (2002) as major predictors of RTI severity and fatality using the multivariate logistic regression models [6,7]. Yau (2004) used the stepwise logistic regression to investigate human and environmental factors associated with severe injuries and demonstrated that drivers' gender is a predicting human factor while road lighting conditions, geographic district, time of crash, and vehicle age are environmental factor predictors of injury severity outcome [8]. Al-Ghamdi et al. (2002) adopted a logistic regression model to identify the growing risk of fatal injuries associated with road types, such as non-intersection roads compared to intersection roads [9]. The study further explored the use of regression models to examine all variables that significantly contribute to road crash fatalities including speed, running a red light, following too close, wrong way, and failure to yield. Tay (2011) presented a multinomial logit model approach to assess pedestrian–vehicle crash severity [10]. Abdel-Aty (2004) compared various models to predict crash injury severity levels and concluded that driver gender, excess speed, use of a seatbelt, vehicle type, and rural or urban crash location are major factors affecting the crash injury severity level [11]. Abdel-Aty (2003) and Xie et al. (2009) generated ordered probit models and Bayesian ordered probit (BOP) models to analyze the driver's injury severity in road crashes [12,13]. El Tayeb et al. (2015) applied association rules' mining algorithms to examine the effect of multiple crash factors on the severity and fatalities of the road in Dubai city, and concluded that male drivers, private vehicles, and the month of December are key factors that highly predicted crash outcomes [14]. Bigham (2014) adopted data mining algorithms, such as logistic regression and classification and regression tree (CART), to underline the influence of human factors on escalating the severity of road crashes [15]. Results revealed the effect of the driver's license possession, seat belt use, gender and age on the severity of road injuries. Several studies carried out in Turkey by Aci (2018) and Akgungor et al. (2009) designed and developed severity models using machine learning methods and artificial neural network models to estimate the number of road injuries and to assess their severe outcomes; they demonstrated that 'cloudiness' and 'high volume of traffic' were major predictor factors that influence the increased occurrence and severity of road crashes in cities [16,17].

Nevertheless, most regression models have inherited limitations, mainly in the assumption of linear and nonlinear relationships between the exploratory variables and the variable under investigation. Violation of any of these presumed assumptions may deviate the analysis and risk potential analysis errors. The overarching objective of this study was to propose a model that will help us to gain a deep understanding of key variables that significantly contribute to fatal road injuries. The aim of this study was two-fold: First, to develop a machine learning-based intelligent model that can accurately classify and rank variables influencing the occurrence of fatal crashes; and secondly, to adopt the proposed machine learning model to investigate the relationship of fatal road crashes with a set of input feature variables. Using the proposed model, the ranking of the input feature variables was identified based on their strong association with the occurrence of fatal road injuries. Gained knowledge from the model outcome can be used to advance our understanding and awareness of major contributing factors associated with fatal crashes and serves as a first step toward addressing these critical elements and improving road safety.

**2. Data Description**

The primary source of data used in this study was the national road crash database procured from the Lebanese Road Accidents Platform (LRAP). The LRAP database compiles national traffic data by crowdsourcing reported road crashes in Lebanon from social media consolidated mainly from three credible sources: Traffic Management Authority, Civil Defense, and Lebanese Red Cross, intending to study crash characteristics and contributing factors [18]. The database encompasses 8482 crash records spanning over a 4-year period from February 2015 to February 2019. As the objective of our study was to identify the variables that significantly contribute to fatal road injuries, data was refined and parsed for the needed information. Nine variables were selected to be included as input variables in this study and one variable 'fatality occurrence' was selected as the output variables (fatal, non-fatal). We obtained the data in digital form and coded the attributes describing the details and outcome of each occurring road crash, including the crash date and time (i.e., month, weekday, hour), location, type (i.e., vehicle, motorcycle, truck, bike, pedestrian), road type (i.e., motorway, primary, secondary, tertiary), injury severity level (no apparent injury, minor injury, serious injury), and the number of fatalities. To account for the spatial feature of the crash data, K-means clustering was applied to all crash events and the cluster ID was recorded as an input variable, 'Spatial Cluster ID', with values ranging from 1 to 10. Table 1 demonstrates the input and output features selected along with their corresponding ranges. Roads' characteristics were extracted from the Lebanese roads shapefile available at Open Street Map.

**Table 1.** Input and output variables.

| Variable | Range |
| --- | --- |
| **Input Variables** | |
| Month | 1–12 |
| Day | 1–31 |
| Day of the Week | Monday–Sunday |
| Hour of Crash | 0–23 |
| AM/PM | am, pm |
| Crash Type | Vehicle–Vehicle, Vehicle–Truck, Vehicle–Pedestrian, Vehicle–Motorcycle, Vehicle–Barrier, Truck–Truck, Truck–Motorcycle, Truck–Barrier, Motorcycle–Motorcycle, Motorcycle–Barrier, Other |
| Injury Severity Level | No Apparent-Injury, Minor Injury, Serious Injury |
| Road Type | Motorway, Trunk, Primary, Secondary, Tertiary, Unclassified |
| Spatial Cluster ID | 1–10 |
| **Output Variable** | |
| Fatality occurrence | Fatal, Not Fatal |

## 3. Method

We designed and developed a machine learning-based model to determine the most influential factors associated with the increased risk of fatal injury occurrence.

*3.1. Model Description*

We developed the model using a classifier ensemble technique. Classifier ensemble is a combination of various individual classifiers that serve to jointly perform the classification task. In the case of diverse individual classifiers (i.e., classifiers disagreeing with each other), their random errors will cancel each other out and will help to yield correct output decisions [19]. We built the model based on the hybrid ensemble technique, which combines homogenous and heterogeneous ensembles of classifiers. A homogeneous ensemble includes classifiers of the same type, while a heterogeneous ensemble, as its name indicates, incorporates classifiers of various types. Adopting homogeneous ensembles may not be the optimal solution, particularly for problems where the ideal base classifier is unclear. Hence, the use of a heterogeneous ensemble is recommended to generate a better performance [20]. Bagging and boosting are commonly used homogeneous ensemble techniques. On the one hand, the bagging technique depicts multiple classifiers that are trained on different under-sampled subsets while allowing them to vote on a final decision. Random forest is an extension of the bagging technique for decision trees. On the other hand, the boosting technique refers to a series of classifiers that are trained on the dataset, with a gradual emphasis on training records that failed to be modelled properly by the previous classifier. Boosting is considered an ultimate approach that reduces the chances of underfitting.

Voting is a heterogeneous ensemble that functions by creating two or more sub-models; each sub-model generates predictions based on a combination of multiple approaches. Upon calculating the mean or the mode of these predictions, each sub-model tends to vote on the ultimate outcome that opts to be adopted. The stacking technique is a simple extension of voting ensembles; it dedicates a model to learn how to best combine the predictions from the sub-models. A meta-model is adopted to best combine the predictions of sub-models. Hence, this technique is often referred to as blending, due to its approach of blending various predictions together.

In this study, we adopted a hybrid combination of heterogeneous and homogeneous ensembles through the ensemble of voting sequential minimal optimization (SMO) with bagging of J48 decision trees. Voting SMO is performed using averaging probabilities, where each base learner produces a classification with a certain probability. The produced classifications are then averaged out, and a new output classification is generated. Bagging of the J48 decision tree is typically adopted as decision trees highly rely upon random node splitting. Details about the chosen model are provided in Section 4.1.

*3.2. Data Pre-Processing*

Prior to using the LRAP data for modelling, we performed a data pre-processing phase to prepare the LRAP data for the model learning and testing phase. In machine learning-based classification problems, data are typically preprocessed to eliminate any imbalances that might negatively affect model performance. Similar to any crash databases, our original LRAP data is inherently biased and imbalanced; the distribution of the two fatality classes (i.e., fatal, non-fatal outcome) was not equally represented in the dataset, with almost 95% of the samples labelled as 'non-fatal road crashes' (ratio 1:19) while only 5% are classified as 'fatal road crashes'. To deal with imbalanced data and to prepare the data prior to its use for model design and development, one of the following two main approaches can be adopted: (i) Use of cost-sensitive learning algorithms or (ii) resampling of the dataset. The latter is a more popular approach due mainly to the lack of cost-sensitive implementations in many learning algorithms [21]. Using appropriate sampling techniques is critical to generating a representative sample dataset. The sampling technique is a preprocessing phase that should be applied to the training set, with the main condition of preserving the validation and test datasets. Resampling the data is performed by either under-sampling of the majority class or by oversampling of the minority class. On the one hand, under-sampling produces a random subset of the majority class. The main drawback of this sampling technique is its potential loss of essential data and relevant information inherent in the original dataset. Oversampling, on the other hand, can simply be achieved through the production of random copies to oversample the minority class, which is referred to as the 'oversampling technique by replacement'. The main disadvantage of the oversampling technique by replacement is the increased risk of data over-fitting as a result of generating replicas of existing instances.

Each sampling technique has its strengths and limitations. While the under-sampling technique poses the potential risk of losing information, oversampling increases the sample size and worsens the needed computational power. To build upon the strengths of the over- and under-sampling techniques, we adopted a hybrid technique called the synthetic minority oversampling technique (SMOTE) as an ultimate technique to achieve a better classifier performance. The SMOTE technique, initially proposed by Chawla et al. (2012), combines the under-sampling and synthetic oversampling approaches to enhance the model's performance. Synthetic oversampling is an oversampling approach in which the minority class is oversampled by creating synthetic examples rather than by replacement [22]. We further adopted the resampling technique on the LRAP data by oversampling the minority class 100% using the SMOTE technique for up to five of the nearest neighbors, followed by the under-sampling technique for the majority class. The new distribution encompasses 83% majority classes and 17% minority classes. This split reduces data imbalance while preserving the proper distribution of data.

To develop the model, we divided the LRAP data into three subsets as follows: 20% of the data was used for testing, and the remaining 80% was split into training and validation subsets. The test dataset was data hidden from the model and used to assess the final performance of a fully developed classifier. The training dataset was used to develop the machine-learning algorithm and the validation data set was used to estimate the model performance while tuning the model's hyperparameters. Validation was conducted through the 10-fold cross-validation technique [23]. The strength of this method resides in its ability to use all observations for both training and validation; each observation was used for validation only once.

*3.3. Metrics Evaluation*

To measure and evaluate the performance of the suggested machine learning-based model, we adopted several assessment metrics from the existing literature. Firstly, the performance of machine learning algorithms is typically evaluated using the predictive accuracy metric shown in Equation (1):

$$Accuracy = \frac{TP+TN}{TP+TN+ FP}, \quad (1)$$

where TP stands for true positive instances, TN for true negative, FP for false positive instances, and FN for false negative instances. Accuracy is a basic metric that might be misleading, and hence more complex metrics are used in the international literature, including the precision (Equation (2)), recall (Equation (3)), and F1-score (Equation (4)) as a weighted average of precision and recall:

$$Precision = \frac{TP}{TP+ FP}, \quad (2)$$

$$Recall = \frac{TP}{TP+FN}, \quad (3)$$

$$F1-score = \frac{2*Precision*Recall}{Precision+ Recall}. \quad (4)$$

The area under the receiver operating characteristic curve (also known as AUROC) shows the balanced classification ability between the true positive rate and false positive rate, though it lacks a suitable metric for datasets with highly imbalanced variables [24]. Nevertheless, the area under precision/recall curve (also known as AUC-PR) is particularly suited for imbalanced datasets in which one class is observed more frequently than the other class. The AUC-PR curves represent a useful alternative to AUROC, particularly when dealing with imbalanced or skewed data sets [25,26].

Finally, we used Cohen's Kappa statistic (Equation (5)) to evaluate classifiers with imbalanced data. Kappa is a measure of how instances, classified by the machine learning classifier, closely matc the data labelled as ground truth, controlling for the accuracy of a random classifier as measured by the expected accuracy. The value of Cohen's Kappa metric ranges from −1 to 1. Landis and Koch (1977) evaluated the performance of a classifier from the value of the Kappa statistic and accordingly classified Kappa values as follows: [0–0.20] as "slight", [0.21–0.40] as "fair", [0.41–0.60] as "moderate", [0.61–0.80] as "substantial", and [0.81–1] as "almost perfect" [27]:

$$Kappa = 1 - \frac{1 - p_o}{1 - p_e}, \quad (5)$$

where $p_o$ is the observed agreement and $p_e$ is the expected agreement.

Within the scope of this study, we used the F1-score, AUC-PR, and Cohen's Kappa statistic as evaluation metrics for our classifier. The modelling was carried out using the Weka environment. Weka is a well-known open-source toolkit that provides a set of machine learning and pre-processing algorithms [28].

## 4. Results

*4.1. Model Development*

The development of the model encompassed two iterations. Initially, as part of iteration 1, we built five different models with a single classifier each. Next, in the second iteration, we leveraged the power of the ensemble technique to achieve a higher performance using the best two classifiers selected from iteration 1.

Finding the best performing classifier was an iterative process. In the first iteration, we constructed a training model using a single classifier. The following five widely adopted learning algorithms were selected to take part in this iteration 1:

1. Sequential minimal optimization (SMO).
2. Random forest.
3. Artificial neural network (ANN).
4. Logistic regression.
5. Naïve Bayes.

For each of these five candidate algorithms, we trained a model to classify the fatality occurrence using the pre-processed dataset, while tuning their hyperparameters to achieve the best performance. The metrics discussed in Section 3.3 were used to measure the performance of each developed model. Table 2 shows the performance metrics for the five classifiers developed during the first iteration. The best performance was achieved by SMO, followed by random forest.

**Table 2.** Performance metrics of single learning classifiers.

|  | F1-Score | AUC-PR | Kappa |
|---|---|---|---|
| SMO | 0.493 | 0.276 | 0.4678 |
| Random Forest | 0.453 | 0.376 | 0.4258 |
| ANN | 0.385 | 0.291 | 0.3462 |
| Logistic Regression | 0.455 | 0.361 | 0.4309 |
| Naïve Bayes | 0.313 | 0.337 | 0.294 |

The ensemble learning discussed in Section 3.1 provides an enhanced classification performance. In the second iteration, we tested various ensemble methods for the two best classifiers: SMO and random forest. SMO is an improved training algorithm based on support vector machines (SVMs). SMO has been proven to provide a huge performance in terms of the needed computation time. As its name infers, random forest selects a random subset of the total features. Out of the selected subset, the best split feature is used to split each node in a tree. Random forest, in its very basic essence, involves building several decision trees. However, one of the disadvantages of using the random forest classifier in Weka is its lack of flexibility for tuning hyperparameters. For this reason, we relied on bagging J48 decision trees, which provide better flexibility. The J48 decision trees algorithm is 'a landmark decision tree algorithm, the most widely used machine learning workhorse in practice to date' [29]. With the bagging of J48 decision trees, all features are considered for splitting a node. This is reflected in Table 3, where the bagging of J48 decision trees 100 times outperforms the random forest of 100 trees.

**Table 3.** Performance metrics using the ensemble method.

|  | F1-Score | AUC-PR | Kappa |
|---|---|---|---|
| Bagging J48 100 decision trees | 0.464 | 0.382 | 0.4365 |
| Vote SMO with Bagging J48 | 0.511 | 0.402 | 0.4882 |

Finally, we identified our best classifier by adopting a hybrid ensemble technique: Vote SMO with bagging J48 as shown in Table 3.

*4.2. Model Performance*

We used the selected classifiers to test the performance of our suggested model. Table 4 shows the classifier performance outcomes on the unseen test set results. Those results are expected to be lower than the training and validation results previously reported in Table 3. A Kappa statistic of 0.4067 shows that the model is considered a moderate classifier.

Table 4. Performance metrics for the selected classifier over the test dataset.

| | Vote SMO with Bagging J48 |
|---|---|
| F1-score | 0.435 |
| AUC-PR | 0.368 |
| Kappa | 0.4067 |

An AUC-PR value of 0.368 is acceptable for our data that have a highly skewed class distribution since there are areas of the precision–recall curve that are unachievable [30]. Comparing the AUC-PR value of 0.386 achieved in our model with imbalanced data of ratio 1:19 (as previously discussed) to the reference AUC-PR value of 0.383 with 1:10 skewed data discussed in [30], we can conclude that our model has satisfactory and effective early retrieval and thus is considered a well-performing model. In Figure 1, the x-axis represents Recall (defined in Eq. 3 above) while the y-axis represents Precision (defined in Eq. 2). The color represents the discrimination threshold value used to get each pair (instance) of Recall/Precision point. In other words, Precision-Recall curve is a parametric curve of the discrimination threshold. Each instance in the curve has a threshold value of class. If the instance highly belongs to this class, its threshold will be 1 so it will have orange color.

Given that our data set is skewed and heavily imbalanced, AUC-PR is the most important metric to evaluate model performance since it focuses mainly on the positive class [31]. Thus, we closely analyzed the obtained AUC-PR results. The baseline of the precision–recall curve can be determined by the ratio of positives (P) and negatives (N) as defined in Equation (6). For our proposed model, the baseline $y$ is equal to 0.076. Most of the precision–recall curve shown in Figure 1 is positioned above the baseline (0.076), which indicates the satisfactory performance of the proposed model according to [31]:

$$y = \frac{P}{P+N}. \tag{6}$$

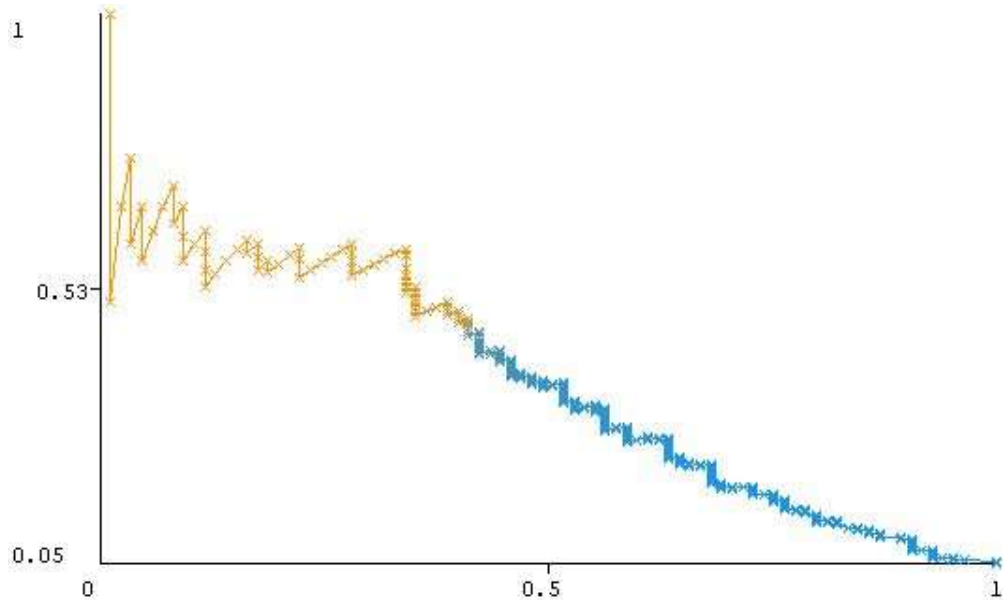

Figure 1. Precision–recall curve for the "Vote SMO with Bagging J48" classifier.

In Figure 1, the x-axis represents Recall (defined in Eq. 3 above) while the y-axis represents Precision (defined in Eq. 2). The color represents the discrimination threshold value used to get each pair (instance) of Recall/Precision point. In other words, Precision-Recall curve is a parametric curve of the discrimination threshold. Each instance in the curve has a threshold value of class. If the instance highly belongs to this class, its threshold will be 1 so it will have orange color.

*4.3. Attribute Evaluation Analysis*

To examine the impact of each input feature and its effect on predicting the output variable (i.e., fatality occurrence), we adopted the attribute evaluator technique within the context of our proposed model. The rank-based attribute selection method assorts the variables based on their importance in predicting the fatality occurrence. The worth of an attribute is calculated by computing the value of the Chi-squared statistic with respect to the class. We selected the Chi-squared statistic as it matches the categorical nature of the variables and applied a 10-fold cross-validation to lower the variance in the results. We compared the Chi-squared values to a critical threshold Chi-squared value calculated based on the assumption of independent variables. A large estimate exceeding the threshold value suggests that the input and the output variables are highly correlated. Estimates below the critical value imply the absence of correlation between the variables. We performed a sensitivity analysis to determine the ranking of the input variables based on their contribution and strong relationship with the fatality occurrence. Sensitivity analysis is a technique widely used to evaluate the effect of input variables' uncertainty on the output variable as well as to assess the robustness of the proposed model. The findings from the sensitivity analysis show that seven out of the nine input variables were correlated with the output variable. Figure 2 depicts the variables ranked in increasing order from the most correlated variable to the least one along with their corresponding Chi-squared values. The impact of some input variables, such as crash type, injury severity level, and spatial cluster ID, is substantial on the fatality occurrence, as depicted in Figure 2. Whereas, other variables, such as month and day of the week, appear to be independent of the increased likelihood of fatal injury occurrence.

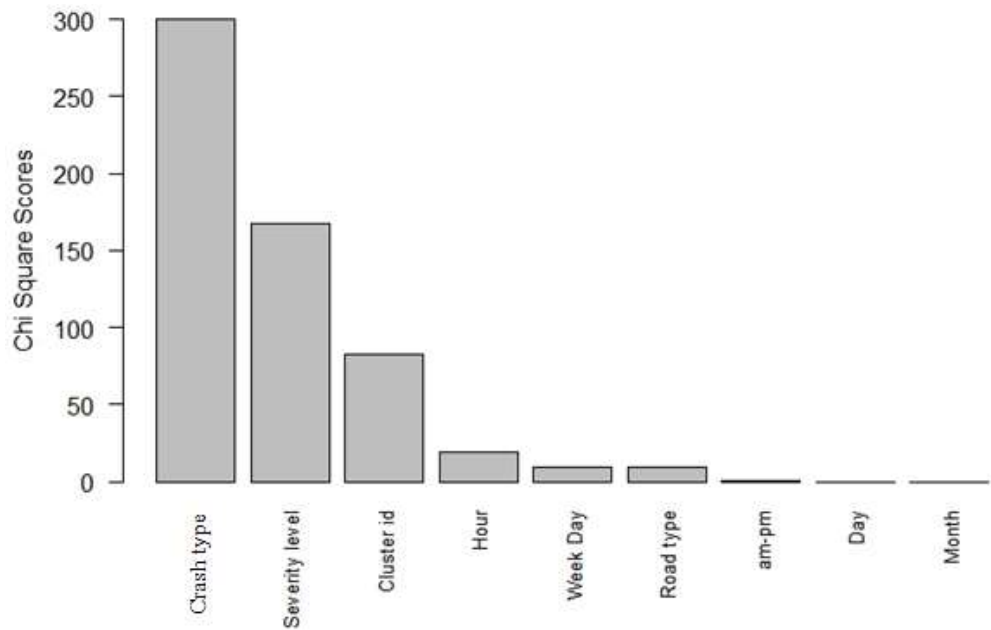

**Figure 2.** Top correlated variables in order of the Chi-squared score.

Using the learning and testing data, our model demonstrated a successful overall prediction of the major contributing factors that increase an individual's likelihood of sustaining a fatal road injury. According to Figure 2, the most influencing variables associated with the increased likelihood of crash fatality are listed in an ascending ranking:

1. Crash type: 'Vehicle–pedestrian' type was the strongest predictor of fatal road crashes. 'Truck–bike' type was the second strongest predictor of fatal road crashes within the crash type category.
2. Injury severity level: The severity level variable was a major contributor to fatal crashes.
3. Spatial cluster ID: Densely populated areas and the presence of major highway crossings proved to be correlated with increased traffic fatalities.
4. Hour of road crash: Time of the day, more specifically 3 am, was highly correlated with fatality occurrence.

5. Day of the week. Friday and Sunday were the two strong factors affecting the occurrence of fatal road injuries.
6. Road type: Motorways are the main road types correlated with fatal crashes.
7. AM-PM and day and month of the year were the least influencing factors on the fatality of the road crash.

**5. Discussion**

The increase in the frequency and severity of road crashes necessitates the effective analysis and abstraction of relevant risk factors associated with fatal road injuries. This study adopted a unique methodological approach to propose a model that synthesizes the strengths of multiple data mining techniques in order to perform a robust analysis of the multiple factors associated with fatal road injuries. Data mining presents a plausible analytical technique widely adopted in scientific analysis; It integrates techniques from multiple disciplines, including database technology, statistics, machine learning, and spatial data analysis, to provide a rich analysis and to allow the extraction of valuable knowledge embedded in datasets. Identifying key fatal injury-contributing factors constitutes the base for informing road safety policies and designing effective preventive countermeasures.

The study findings align with existing studies and reveal that 'crash type', mainly the 'vehicle–pedestrian' and the 'truck–motorcycle' types were the most significant factors associated with the increased risk of road fatalities. This is evident as pedestrians and motorcyclists are particularly vulnerable road users [32]. Studies in many low- and middle-income countries (LMICs) indicate that pedestrians disproportionally account for the majority of road fatalities [1], particularly that a large number of individuals commute by walking- the most affordable mode of transportation. With lax safety policies and the absence of road infrastructure and regulations (e.g., designated and lighted crosswalks, road safety design, and traffic management), many pedestrians are forced to share the roads with vehicles, which places them at a greater risk of being hit or killed on the road. Many LMICs reported high rates of pedestrians' fatalities, reaching up to 85% of total road victims [33–37]. This is mainly due to pedestrians' low tolerance to high impact vehicle collisions, often resulting in severe bodily damage and fatal injuries. In some instances, pedestrians' unsafe behaviors tend to exacerbate the risk of fatal injuries. According to the Committee of Land Transport in the Global Safety Organization [38], safety behaviors, namely the lack of compliance with existing road safety regulations and laws (e.g., ignoring traffic signals, crossing midblock), increased pedestrians' risk of injuries. Concerted efforts from multiple entities, including road engineers, law enforcement officers, and non-governmental agencies, should be mobilized to encourage safe behaviors and increase pedestrians' safety on the roads.

The 'truck–motorcycle type' was the second strongest predictor of fatal road crashes within the 'crash type' category. Given the disproportional size difference between the two transport modes, the risk of fatal injury substantially increases when a truck–motorcycle crash occurs. Moreover, the lack of motorcycling infrastructure on many roads imposes a great risk on motorcycle riders [39] as motorcycles are often not visible to truck drivers, particularly on narrow shared roads, at intersections, or when the motorcycle is caught in the truck's blind spot. Another major contributor to fatal injuries is when trucks attempt to turn left while motorcycle riders keep travelling straight, leading to high impact collisions and fatal injury [40]. The implementation of sensors or assistant systems that alert truck drivers of dangerous situations or warn them of blind spot occupancy can substantially prevent fatal collisions. Moreover, wearing protective clothing, using helmets, and abiding with road safety regulations are measures that considerably reduce the fatality of the motorcyclists on the road.

Our analysis demonstrated that the injury severity level is a critical factor in determining the fatality outcome of the road injury. Individuals involved in road crashes with high injury severity levels are strongly correlated with death outcomes; road victims with reportedly high injury severity scores (i.e., ISS 51–75) suffer from low survival rates compared to victims with a low ISS score of 40 or less [41]. Several studies confirmed the strong association between fatal road crashes and severe head trauma, particularly among pedestrians and motorcyclists [32,42]. Head injuries are considered a severe injury, leading to 4–5 times higher risk of road deaths. The reduced fatality of high ISS is mostly attributed to the enhanced medical services provided at the crash site, coupled with advancement in medical technologies and treatment at local trauma centers and medical facilities. Future work should focus on improving regional medical services and accessing these services through mapping fatal road crash sites relative to local emergency centers' locations.

Geographic location and environmental factors were also highly associated with fatal crashes [43]. This study shows that 'spatial cluster ID' was another major contributor to fatal injuries. Fatal crashes mainly clustered in densely populated residential locations and in areas of high socioeconomic status families. This clearly indicates that areas populated with families of a high socioeconomic status tend to observe a high number of privately owned vehicles and consequently an excess volume of traffic. With the increased number of private vehicles possessed per household and less reliance on public transport, the vehicle miles traveled (VMT) increases and ultimately the probability of fatal road crashes increases.

Another predictor of road fatality was the 'hour of road crash' factor. This finding agrees with existing literature and confirms the high correlation between early AM crash time and fatality outcome [7,37]. Crashes mostly happening between 1 and 6 AM result in a disproportionally high rate of death compared to crashes occurring during the day [7,37]. According to a study conducted in the United Kingdom, injury severity, defined as the number of deadly crashes per 100 crashes, is higher during nighttime than daytime [44]. The heightened fatal crashes in the early AM hours are mainly associated with multiple factors, including drivers' fatigue and sleepiness, alcohol consumption, and reduced visibility at night, particularly on non-lit roads. Sleepiness and fatigue are critical conditions that result in drivers' diminished cognitive performance and proper risk judgment, especially following a long working day [45]. Advanced technologies and sensors should be embedded in vehicles to help detect the driver's sleepiness mode and alert the vehicle's occupant. Impaired driving regulations should be strictly enforced by government police. Placing radar detectors and performing alcohol tests on drivers at late hours would help to mitigate the occurrence of fatal injuries. Clear visibility is a major protective factor; the absence of a properly lit road hinders drivers' clear visibility at night and consequently reduces their information processing capabilities and slows their reaction time during crashes [44], which translates into longer stopping distances, and increases the severity of crash outcomes.

The result of the analyses further showed that crashes occurring on a specific day of the week are more correlated with fatal outcomes. According to our study, Fridays and Sundays are associated with an increased probability of fatal crashes. Fridays and Sundays correspond to the beginning and end of weekends in Lebanon, with increased leisure travel and excessive traffic commuting and exiting main urban cities. Ample studies reported high rates of fatal road injuries occurring on weekends; approximately 38% to 56% of fatal road crashes occur on Fridays to Sundays [34,46,47]. Typically, fatal road injuries are more prevalent on weekends, major holidays, and Fridays leading to long weekends [48]. To curtail the number of fatal road crashes, strong enforcement should be rigorously implemented to enhance safety and reduce fatalities on the road.

Another main contributing factor for fatal road crashes was 'road type'. Compared to urban and rural roads, highways were specifically correlated with fatal crashes. A number of studies demonstrated that fatal injuries mostly occur on intercity roads and highways outside of the urban setting, leading to an increased frequency and fatalities of road crashes [7,42,49]. This increased crash fatalities on highways is due to multiple reasons, mainly road infrastructure, speed limit variation, and road lighting conditions. The lack of safety in road design and infrastructure presents additional potential hazards for severe injuries, including crashes occurring at sharp curves, non-lit intersections, and slippery pavement surfaces [50].

This demonstrative model has many strengths and limitations. One of the major advantages offered by this model is its ability to predict the occurrence of fatal road crashes and to identify the interplay of several contributing elements that increase the likelihood of these fatal crashes. An advanced pre-processing technique was first used to account for the dataset's skewness. The model was developed based on hybrid ensemble machine learning, which proved to be highly accurate. This model encountered some limitations. A limited number of variables were not available in the dataset, such as sex and alcohol blood level; therefore, an interpretation of the data predictions should be considered in light of the many assumptions and the availability of variables in the dataset. Future research questions remain to be answered, particularly concerning how sensitive the classification is to crash parameters.

## 6. Conclusions

This study's main contribution lies in the proposed methodology that provides insight into the design, development, and validation of a demonstrative machine learning-based model in assessing the relevance and influence of multiple contributing factors on the desired outcome, adding value to the literature of data mining, with applicability to various case studies in many fields and disciplines, including road safety and injury prevention. The proposed model employed a hybrid ensemble classification technique that combines voting SMO and the bagging of J48 decision trees using the LRAP road crash database of 8482 cases. Through validation, the proposed model proved to be an effective analytical model that can be adapted to provide effective overall predictions of the multiple factors associated with fatal crash occurrence. Findings from this study can be implemented by road safety stakeholders and law enforcement to design tailored and injury prevention-specific regulations and safety countermeasures taking into consideration these major contributors to road fatalities. Further research warrants the analysis of additional road injury crashes to enhance the model's prediction performance and the accuracy of its estimated results, as well as to reveal additional factors that contribute to the increased risk of sustaining fatal road injuries.

**Funding:** This research was funded with support from National Council for Scientific Research in Lebanon.

**Conflicts of Interest:** The authors declare no conflict of interest.